\documentclass[aps,pra,superscriptaddress,twocolumn]{revtex4-1}
\usepackage{times}
\usepackage[utf8]{inputenc}
\usepackage[T1]{fontenc}
\usepackage{amsmath}
\usepackage{amssymb}
\usepackage{graphicx}
\usepackage{tikz,pgfplots}
\usepgfplotslibrary{fillbetween}
\usetikzlibrary{backgrounds}
\usetikzlibrary{patterns}
\usepackage{hyperref}
\usepackage{array}
\usepackage{mathrsfs}
\usepackage{color}
\usepackage{soul,xcolor}
\setstcolor{red}
\newcommand{\be}{\begin{equation}}
\newcommand{\ee}{\end{equation}}
\newcommand{\br}{\begin{eqnarray}}
\newcommand{\er}{\end{eqnarray}}

\begin{document}
\title{Temperature estimation of an entangled pair of trapped ions}
\author{O. P. de S\'{a} Neto}
\affiliation{Coordena\c{c}\~ao de Ci\^encia da Computa\c{c}\~{a}o, Universidade Estadual do Piau\'{i}, 64202--220, Parna\'{\i}ba, Piau\'i, Brazil.}
\author{H. A. S. Costa}
\affiliation{Departamento de F\'{\i}sica, Universidade Federal do Piau\'{\i}, 64049-550, Teresina, Piau\'i, Brazil.}
\author{G. A. Prataviera}
\affiliation{Departamento de Administra\c{c}\~ao, FEA-RP, Universidade de S\~{a}o Paulo, 14040-905, Ribeir\~{a}o Preto, SP, Brazil}
\author{M. C. de Oliveira}
\affiliation{Instituto de F\'{\i}sica Gleb Wataghin, Universidade Estadual de Campinas, 13083-970, Campinas, S\~ao Paulo, Brazil.}
\email{marcos@ifi.unicamp.br}


\begin{abstract}
We apply estimation theory to a system formed by two interacting trapped ions. By using the Fisher matrix formalism, we introduce a simple scheme for estimation of the temperature of the longitudinal vibrational modes of the ions.  We use the ions interaction to effectively infer the temperature of the individual ions, by optimising the interaction time evolution and by measuring only over one of the ions. 
We also investigate the effect of a non-thermal reservoir over the inference approach. The non-classicality of one of the ions vibrational modes, introduced due to a squeezed thermal reservoir, improves the indirect inference of the individual temperatures.
\end{abstract}

\maketitle
%
%


\section{Introduction}

 In the quantum mechanical context, the temperature of a system is a nonlinear function of the density operator, so it cannot directly correspond to a quantum observable - one has to indirectly estimate its value by measuring another observable. This indirect procedure for temperature estimation implies in an additional uncertainty for the measured value, which should be handled appropriately. Therefore, any strategy aimed to determine the temperature of a quantum system reduces to a parameter estimation problem \cite{Paris, Paris2, Helstrom}.  Quantum theory of estimation (QET) provides a formal framework in order to optimise the inference procedure by minimising the additional uncertainty \cite{Holevo, Giovannetti}.  

 Techniques of quantum parameter estimation have been  devoted to estimate the temperature in the context of quantum thermodynamics \cite{razavian2019quantum,campbell2017global,farajollahi2018estimation,hofer2017quantum} and also for technological applications in many different branches of Science - ranging from Material Sciences to Biology and Medicine \cite{Yang, Kucsko}. A central purpose is to employ a controlled quantum system (with view to applications in low temperature measurement \cite{DePasquale2018}) to explore the thermodynamics in the regime of small-scale physics \cite{Brunelli, Marzolino, Salvatori, Correa}, where quantum effects become predominant \cite{Allahverdyan, Hilt, Williams}. Several quantum systems have been used to estimate very low temperature such as Bose-Einstein condensates \cite{Sabin}, ultracold lattice gases \cite{Mehboudi}, trapped ions \cite{Turchette, Tolazzi, Ivanov}, single-qubit \cite{razavian2019quantum}, to mention a few.

  In this work, we address the issue of the vibrational degrees of freedom temperature estimation in a system of two interacting trapped ions. Our interest is to investigate how precisely we are allowed to estimate the local differences of temperatures of vibrational degrees of freedom of the two ions simultaneously. To this purpose, we employ the multiparameter Fisher information
 \begin{eqnarray}
 \mathcal{F}_{\alpha\beta} = \sum_{r_i} \mathcal{P}(r_i)\left(\frac{\partial \ln\mathcal{P}(r_i)}{\partial \theta_\alpha}\right)\left(\frac{\partial \ln\mathcal{P}(r_i)}{\partial \theta_\beta}\right), 
 \end{eqnarray}
 to investigate the parameter estimation accuracy \cite{Kay, Liu,ataman2020single}, where $\theta_i$ denotes the outcome of a measurement, $\mathcal{P}(r_i)=p(r_i|\theta_i)$ is the conditional probability of measuring $r_i$ if the value of the parameter under consideration is $\theta_i$. From the practice standpoint, increasing the Fisher information of the system tends to increase the maximum precision that can be obtained by an estimation scheme. Mathematically, this relationship is quantified by the Cram\'er-Rao lower bound \cite{Cramer, Rao}, 
 \begin{eqnarray}
    \mathrm{Var}(\hat{\theta}_i) \geq \frac{1}{\sqrt{\mathcal{F}_{ii}}},
 \end{eqnarray}
 where $\hat{\theta}_i$ is the estimator of the unknown parameter $\theta_i$. The Cram\'er-Rao lower bound is saturated asymptotically by the optimization of the elements of the Fisher information matrix via a suitable choice of all its dependent parameters. Quantum parameter estimation purpose in quantum thermometry is to use a measurement over a quantum probe system, to infer the temperature of the reservoir it is immersed. For that purpose, an optimisation over the measurement operation is usually employed in order to find the maximal overall Fisher information (and consequently the measurement with the minimal dispersion). However, the procedures employed are sometimes cumbersome and quite generally do not bring any information on how and which observable is to be measured \cite{DePasquale2018}. Such a difficulty is even worse in the situation when systems of continuous variables are involved \cite{Correa, safranek1, safranek2}. Here we take a more pragmatical approach, instead of optimising the measurement operator, we employ measurements accessible in actual ion experiments\cite{brown2011coupled, harlander2011trapped} for the calculation of the Fisher information. We employ the detection of one of the ions vibrational mode phonon number, which is accessed experimentally through the observation of the asymmetry between the red and blue motional sidebands of hyperfine Raman transitions\cite{Monroe1, Monroe2}. The phonon number detection on one of the ions (say ion 1) is then employed for simultaneous inference of the temperature of both ions.

 In what follows, firstly we introduce the theoretical model and we show how to describe the ionic vibrational modes. We propose a joint estimation scheme, where a single probe state is used to estimate the temperatures of the two ions ($T_1$ and $T_2$) with a single projective measurement corresponding to the number of excitations in the first ion vibrational degree of freedom. In particular, we calculate the Fisher information as a function of the temperature of the ions and the interaction parameter $g$. We analyse the Fisher information of both ions in two scenarios -- In the first one, we consider ions with distinct temperatures. Secondly, to infer how other changes in the vibrational state of one ion affects the performance of our approach, we consider a non-thermal bath in equilibrium with one of the ions.

\section{Two Interacting Charges}

\begin{figure}
\centering
\includegraphics[scale=0.5]{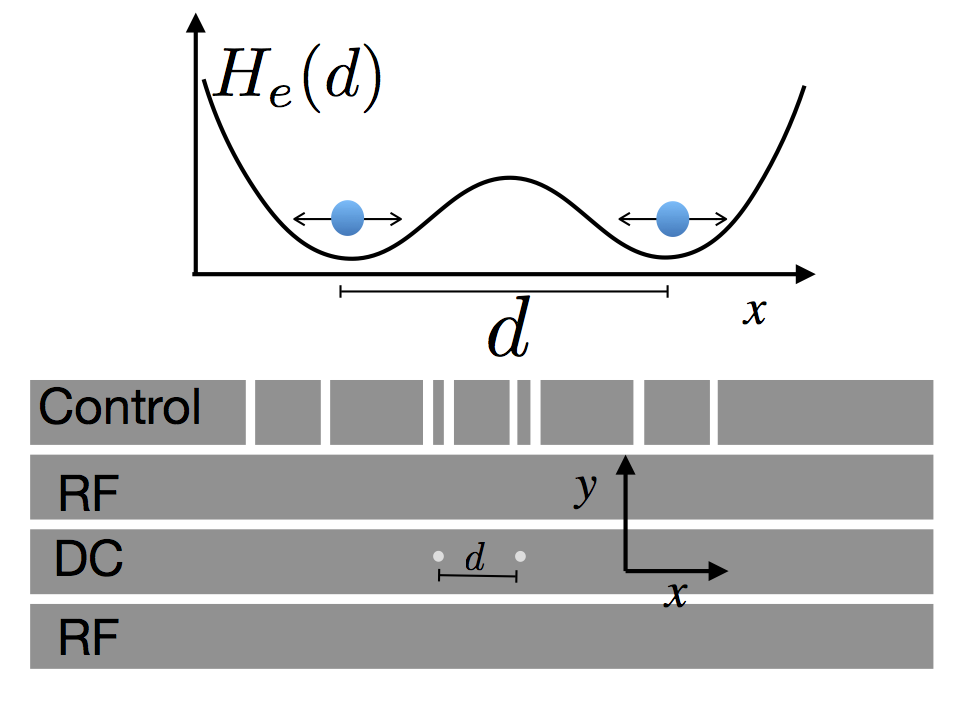}
	\caption{Scheme of two interacting trapped ions with the same charge $q$. Each ion is confined in a potential well along of coordinate $x$ and separated by a distance $d$. Schematic diagram based on Ref. \cite{brown2011coupled}.}
	\label{SM}
\end{figure}

 We consider two interacting trapped ions having the same charge $q$ and mass ${m}$. Each ion is confined in a potential well along of coordinate $x$ and separated by a distance $d$, as shown in Fig. (\ref{SM}), accordingly with refs. \cite{brown2011coupled, harlander2011trapped}
 
Considering only the coordinates along the longitudinal separation of the ions, the Hamiltonian that describes this system reads \cite{brown2011coupled, harlander2011trapped}
\begin{eqnarray}
H = \sum_{j=1}^2\left(\frac{p_{j}^{2}}{2m}+\frac{1}{2}m\omega^{2}x_{j}^{2}\right)+H_{e},
\label{H}
\end{eqnarray} 
where $\omega$ is the frequency of oscillation of each ion due to the trapping potential, and $p_i$ and ${x_i}$ are the momentum and position of the ion $i=1,2$, respectively. The term $H_{e}$ is the ions  electrostatic interaction energy given by \cite{prytz2015electrodynamics}
\begin{eqnarray} \label{Hc}
H_{e}&=&\frac{1}{4\pi\epsilon_{0}}\frac{q^2}{d+ x_2 - x_1  },
\label{Hm}
\end{eqnarray}
where $(4\pi\epsilon_{0})^{-1}=9\times 10^{9} Nm^{2}/C^{2}$.
By expanding (\ref{Hc}) in powers of $\frac{x_{2} - x_{1}}{d}$, and considering small relative displacements we can rewrite the Coulomb interaction energy $H_{e}$ approximately as
\begin{eqnarray}
H_{e}(d) \approx \frac{q^2}{4\pi\epsilon_{0}d}\left[1-\frac{x_{2} - x_{1}}{d}+\frac{\left(x_{2} - x_{1}\right)^{2}}{d^{2}}\right].
\label{hcp}
\end{eqnarray}

Introducing the operators 
\begin{equation}
 x_{i}=\sqrt{\frac{ \hslash}{2m\omega}}(a_{i}^{\dagger} +a_{i}),\hspace{1.cm} i=1,2,
 \end{equation}
 and
 \begin{equation}
 p_{i}=i\sqrt{\frac{ \hslash m\omega }{2}}(a_{i}^{\dagger}-a_{i}),\hspace{1.0cm} i=1,2,
\end{equation}
the total system Hamiltonian becomes
\begin{eqnarray}
 H &=& \hslash \Omega a_{1}^{\dagger}a_{1}+ \hslash \Omega a_{2}^{\dagger}a_{2} +\sqrt{\frac{\hslash k^2 q^4}{2m\omega d^4}}(a_{1}+a_{1}^{\dagger})\nonumber\\
 &&-\sqrt{\frac{\hslash k^2 q^4}{2m\omega d^4}} (a_{2}+a_{2}^{\dagger})+\frac{\hslash kq^2}{2m\omega d^{3}}(a_{1}^{2}+a_{1}^{\dagger 2})\nonumber \\ &&
 +\frac{\hslash k q^2}{2m\omega d^{3}}(a_2^{2}+a_{2}^{\dagger 2})
 \nonumber \\ &&+\frac{\hslash k q^2}{m\omega d^{3}}(a_{1}a_{2}^{\dagger }+a_{2} a_{1}^{\dagger }+a_{1}a_{2}+a_{1}^{\dagger}a_{2}^{\dagger }), 
\end{eqnarray}
where $\Omega=\omega+(kq^2/m\omega d^3)$ are the ions trap frequencies deviated by a small frequency shift due the ions interaction, and $k=1/4\pi \epsilon_{0}$.
In the interaction picture obtained by the unitary transformation $U=e^{iH_{o}t}$, where $H_{o}=\hslash \Omega a_{1}^{\dagger}a_{1}+ \hslash \Omega a_{2}^{\dagger}a_{2}$ corresponds to the trapped ions free evolution, the Hamiltonian becomes 

\begin{widetext}
\begin{eqnarray}
\tilde{H} &=&  \sqrt{\frac{\hslash k^2 q^4}{2m\omega d^4}}\left[(a_{1}e^{i\Omega t}+a_{1}^{\dagger}e^{-i\Omega t}) - (a_{2}e^{i\Omega t}+a_{2}^{\dagger}e^{-i\Omega t})\right]
 \\ && +\frac{\hslash kq^2}{2m\omega d^{3}}\left[(a_{1}^{2}e^{2i\Omega t}+a_{1}^{\dagger 2}e^{-2i\Omega t}) +(a_2^{2}e^{2i\Omega t}+a_{2}^{\dagger 2}e^{-2i\Omega t})\right]\nonumber\\
 &&+\frac{\hslash k q^2}{m\omega d^{3}}(a_{1}a_{2}^{\dagger }+a_{2} a_{1}^{\dagger }+a_{1}a_{2}e^{2i\Omega t}+a_{1}^{\dagger}a_{2}^{\dagger }e^{-2i\Omega t}), 
\end{eqnarray}
\end{widetext}
and the terms oscillating at frequencies $\Omega$ and $2\Omega$ are averaged to zero for typical scales of time and can be disregard. The ions interaction Hamiltonian reduces to the simple form 
\begin{equation}\label{H12}
\tilde{H}=\hslash g \left( a_{1}a_{2}^{\dag}+a_{1}^{\dag}a_{2}\right),
\end{equation}
with the coupling constant $g = kq^{2}/m\omega d^{3}$. This coupling Hamiltonian is similar in form to that one for two radiation modes interacting through a beam-splitter. It is well known that a beam-splitter can only entangle two light modes in at least one of them is non-classical\cite{kim, xiang, DEOLIVEIRA}. Therefore, it is interesting to investigate whether or not the non-classicality of one of the states affects the process of temperature inference. For that, in the next section we develop a formalism that allows to tackle the dynamics of the two ions vibrational modes independently of the initial equilibrium conditions.

\section{Bipartite Gaussian States}

 All states  easily accessed experimentally for our purposes, are Gaussian. Therefore, here we develop the formalism more appropriate to investigate the properties of the ionic vibrational degrees of freedom. A two-mode
bipartite quantum state $\rho$ is Gaussian if its symmetric
characteristic function \cite{de2005characterization} is given by 
$C({\mathbf{\eta}})=Tr[D({\mathbf{\eta}})\rho]=e^{-\frac12{\mathbf{\eta}^\dagger}{\mathbf
V}{\mathbf{\eta}} }$, where
$D(\mathbf{\eta})=e^{-\mathbf{\eta}^\dagger{\mathbf E}{\mathbf v}}$ is a
displacement operator in the four-vector
$\mathbf{\eta}$-space: 
$\mathbf{\eta}^\dagger=\left(\eta_1^*, \eta_1, \eta_2^*,
\eta_2\right)$, ${\mathbf v}^\dagger= \left(a_1^\dagger, a_1,
a_2^\dagger, a_2\right)$, and
  \be \mathbf{
E}=\left(\begin{array}{c c}{\mathbf{Z}}&\mathbf{ 0}\\\mathbf{0}&
{\mathbf{Z}}\end{array}\right),\;\;\;
{\mathbf{Z}}=\left(\begin{array}{c c}{1}&{0}\\{0}&
{-1}\end{array}\right),\ee where $a_1$ ($a_1^\dagger$) and $a_2$
($a_2^\dagger$) are annihilation (creation) operators for party 1
and 2, respectively. ${\mathbf V}$ is a $4\times 4$ covariance matrix
with elements
$V_{ij}=(-1)^{i+j}\langle\{v_i,v_j^\dagger\}\rangle/2$, which can
be decomposed in four block $2\times 2$ matrices, \be\label{var}
{\mathbf V}=\left(\begin{array}{c c}{\mathbf V_1}&{\mathbf C}\\{\mathbf
C}^\dagger& {\mathbf V_2}\end{array}\right),\ee
where ${\mathbf V_1}$ and ${\mathbf V_2}$ are $2\times2$ Hermitian matrices
containing only local elements, while ${\mathbf C}$ is a $2\times2$ matrix representing the correlation
between the two parties and explicitly are written as
\be \mathbf{
V_i}=\left(\begin{array}{c c} n_i&m_i\\m_i^*&
n_i\end{array}\right), i=1,2, \;\;\;
{\mathbf{C}}=\left(\begin{array}{c c}{m_s}&{m_c}\\{m_c^*}&
{m_s^*}\end{array}\right).\ee
Positivity and separability for bipartite
Gaussian quantum states have been largely investigated
\cite{englert,simon,schur}.
 Besides
the requirement of the uncertainty
principle  \be \label{cond}{\mathbf V}+\frac 12
{\mathbf E}\ge 0,\ee  there is a necessary and sufficient condition, which must be satisfied for separable
Gaussian states \cite{schur} \be
\mathbf {\widetilde V}+\frac 12 \mathbf{E}\ge 0,\label{cond2}\ee under a
partial phase space mirror
reflection, 
 $\mathbf{\widetilde V}=\mathbf {TVT}:{\mathbf T \mathbf v^\dagger}={\mathbf v^\dagger}_T=
 \left(a_1^\dagger,
a_1,a_2,a_2^\dagger\right)$, with \be {\mathbf
T}=\left(\begin{array}{c c}{\mathbf I}&{\mathbf 0}\\{\mathbf 0}& {\mathbf
X}\end{array}\right),\, \text{and} \,\, {\mathbf X}=\left(\begin{array}{c
c}{ 0}&{ 1}\\{1}& {0}\end{array}\right).\ee

The physical
positivity criterion (\ref{cond}) applies only if \cite{schur}
 \br \label{sep}n_1&\ge&
\sqrt{|m_1|^2+\frac 1 4},\\ n_2&\ge& \frac{s}{d}+\sqrt{\frac1 4
\left[\frac{\left||m_c|^2-|m_s|^2\right|}{d}-1\right]^2+|m_2-c|^2},\er
respectively,  with $
s=n_1\left(|m_c|^2+|m_s|^2\right)-m_cm_sm_1^*-m_c^*m_s^*m_1$,
$c=2n_1m_s^*m_c-m_c^2m_1^*-(m_s^*)^2m_1$, and $ d=n_1^2-\frac1
4-|m_1|^2$. Similarly
 the separability condition (\ref{cond2}) writes explicitly into (\ref{sep}) and
 \br \label{sep2}n_2&\ge& \frac{s}{d}+\sqrt{\frac1 4
\left[\frac{\left||m_c|^2-|m_s|^2\right|}{d}+1\right]^2+|m_2-c|^2}.\label{entangled}\er
%

The ions vibrational modes are assumed as initially uncoupled, so that in (\ref{var}) $\mathbf{C}=\mathbf{0}$, and they prepared in special local states to be discussed later. However, the effect of the interaction (\ref{H12}) is to correlate the two modes. This can be be seen as the following Bogoliubov operation $B$:
\begin{eqnarray}
\rho_{out}&=&B\rho_{in}B^\dagger,\label{1}\\
B{\mathbf v}B^\dagger
&=&{\mathbf M} {\mathbf v},\label{2}\\
{\mathbf M}&=&\left(\begin{array}{c c}{\mathbf R}&{\mathbf S}\\{\mathbf -S}^*&
{\mathbf R}^*\end{array}\right),
\end{eqnarray}
com
\begin{equation}
 {\mathbf R}=\cos\theta\left(\begin{array}{cc} e^{i\phi_0}&0\\0&
e^{-i\phi_0}\end{array}\right),
{\mathbf
S}=\sin\theta\left(\begin{array}{cc}e^{i\phi_1}&0\\0&e^{-i\phi_1}\end{array}\right),\nonumber
\end{equation} where
$\rho_{out}$ is the density operator for the joint output state.
The output symmetric characteristic function is given by \be
C_{out}({\mathbf{\eta}})=Tr[D({\mathbf{\eta}})\rho_{out}]=Tr[B^\dagger
D({\mathbf{\eta}})B\rho_{in}].\ee Now with the help of Eqs.(11-13),
$ B^\dagger D({\mathbf{\eta}})B=e^{-\mathbf{\eta}^\dagger{\mathbf E}{\mathbf
M}^{-1}{\mathbf v}}\equiv D({\mathbf{\zeta}})$, with $\mathbf{\zeta}={\mathbf
M}\mathbf{\eta}$, since ${\mathbf M} {\mathbf E M}^{-1}={\mathbf E}$. Thus \be
C_{out}({\mathbf{\eta}})=C_{in}({\mathbf{\zeta}})=e^{-\frac12{\mathbf{\zeta}^{\dagger}}{\mathbf
V}{\mathbf{\zeta}} }=e^{-\frac12{\mathbf{\eta}^{\dagger}}{\mathbf
V^\prime}{\mathbf{\eta}} },\label{car}\ee
 where ${\mathbf
V}^\prime={\mathbf M}^{-1}{\mathbf V}{\mathbf M}$, and 
analogously to (\ref{var}), ${\mathbf V}^\prime$ can be block
decomposed
with \br {\mathbf V_1^\prime}&=& {\mathbf R}^{*}{\mathbf V_1}{\mathbf R}+{\mathbf
S}{\mathbf V_2}{\mathbf S}^*,\\
 {\mathbf V_2^\prime}&=& {\mathbf
S}^{*}{\mathbf V_1}{\mathbf S}+{\mathbf R}{\mathbf V_2}{\mathbf R}^{*},\\
 {\mathbf C^\prime}&=& {\mathbf
R}^{*}{\mathbf V_1}{\mathbf S}-{\mathbf S}{\mathbf V_2}{\mathbf R}^{*}.\er 

For the specific unitary transformation induced by the time-indepented two-modes coupling (\ref{H12}) $\phi_0=\phi_1=0$, and $\theta= g t$, where $t$ is the time variable. 

\begin{figure}[ht]
  \includegraphics[scale=0.6]{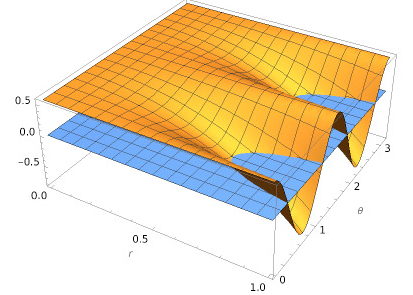}
	\caption{Entanglement of the ions logitudinal vibrational as a function of the squeezing parameter of ion 2, and the interaction $\theta =gt$. Only for $2(\bar{n}_{2}+1/2)\leq e^{2r}$, is that the vibrational mode 2 is non-classical, allowing that the evolution (\ref{ucor}) entangle the ions (represented by the negative values of the vertical axis). Maximal entanglement occurs at $\theta =\pi/4$ (or $3\pi/4$), corresponding to a 50:50 beam-splitter.} 
\label{corr}
\end{figure}

 \section{Fisher Information} 
 
The fluctuations in the trap parameters may introduce perturbations in the equilibrium state of each distinct ion. Typically, the ions coupling parameter $g$ is very small, and therefore the temperature of the system must reach cryostatic regimes of a few kelvins. Even so, in \cite{brown2011coupled}, e.g., a thermal heating rate of 1,885 quanta per second was observed. Therefore there may be an inhomogeneity of the ions equilibrium temperatures. Further experimental limitations constrains the direct access to all ions in the trap \cite{brown2011coupled}, as was suggested by the inference approach in \cite{Ivanov}. Therefore, our purpose here is to use the phonon number detection (over one of the ions only) as a way to infer simultaneously the temperature differences of both ions. 

 As mentioned before, we shall consider the case where the ions are initially uncorrelated so the total density operator factorises as
\begin{eqnarray} \label{rho12}
  \rho_{12} = \rho_{1}\otimes\rho_{2},
\end{eqnarray}
where $\rho_1$ and $\rho_2$  are the reduced density operators of ions 1 and 2, respectively. We assume that the ion 1 is initially in equilibrium with a thermal reservoir at temperature $T_1$, while ion 2 can be in equilibrium  with a thermal or non-thermal (for future purposes) reservoir at temperature $T_2$. For that, we introduce a controllable squeezing parameter so that the ion 2 can be prepared in a thermal squeezed state, respectively, by
\begin{equation}
 \rho_1 = \int d^{2}\alpha \frac{e^{-\frac{|\alpha|^2}{\bar{n}_1}}}{\pi \bar{n}_1}
 | \alpha \rangle\langle \alpha |, \;\;
 \rho_2 = \int d^{2}\beta \frac{e^{-\frac{|\beta|^2}{\bar{n}_2}}}{\pi \bar{n}_2}
S | \beta \rangle\langle \beta | S^{\dag},
\end{equation}
where
\begin{eqnarray*}
S &=& e^{\frac{r}{2}\left(a_{2}^{2}+a_{2}^{\dag 2}\right)},
\end{eqnarray*}
is the one-mode squeezing operator, and the average thermal photon number given by
\begin{eqnarray}
\bar{n}_1 = \frac{1}{e^{\frac{\hslash\Omega}{k_{B}T_1}} - 1},\quad \bar{n}_2 = \frac{1}{e^{\frac{\hslash\Omega}{k_{B}T_2}} - 1},
\end{eqnarray}
 where $T_1$ and $T_2$ correspond to the equilibrium temperature for the ion-1 and ion-2, respectively.

 The time evolution of the density operator with Hamiltonian (\ref{H12}) occurs with the following expression
\begin{eqnarray} \label{rho12t}
 \rho_{12}(\theta) = U(\theta)\rho_{12}U^{\dagger}(\theta),
\end{eqnarray}
where ($\theta=gt$)
\begin{eqnarray}\label{ucor}
   U(\theta) = \exp[-i\theta(a_{1}a_{2}^{\dag} + a_{1}^{\dag}a_{2})].
\end{eqnarray}    

To understand the correlation induced by (\ref{ucor}), we plot in Fig. \ref{corr} the bound (\ref{entangled}) as a function of $\theta$ and the squeezing parameter $r$ for the ion 2 for a fixed temperature of $T_{1}=T_2=2.8\times10^{-5}$ K. We see that only for $2(\bar{n}_{2}+1/2)\leq e^{2r}$, is that the vibrational mode 2 is non-classical, allowing that the evolution (\ref{ucor}) entangle the ions. as it is expected the maximal entanglement occurs at $\theta =\pi/4$ (or $3\pi/4$), which corresponds to a 50:50 beam-splitter.
   
\begin{figure}[ht]
   \includegraphics[scale=0.33]{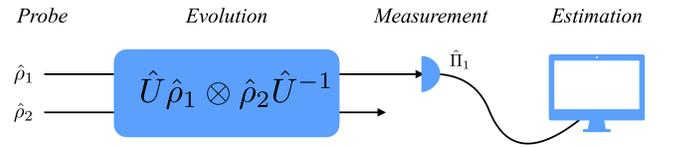}
	\caption{ Scheme for the indirect estimation of the temperatures of two trapped ions interacting via an electromagnetic interaction. After some interaction time between the two ions, a energy measurement  corresponding to the number of particles is performed on the 1-ion, followed by post-processing for the estimation of parameters $T_1$ and $T_2$.} 
\label{Illustration}
\end{figure}

 By applying estimation techniques to the state $\hat{\rho}_{12}(t)$ we can achieve the ultimate bound of parameter $T_1$ or $T_2$ using measurement schemes feasible with current technology. In particular, the estimation of the temperatures of the ions can be implemented by choosing the projective measurement corresponding to the number of phonons  in the longitudinal vibrational mode mode of the ion 1 , as described in Fig. \ref{Illustration}. It shows our estimation scheme, where the dynamics of two trapped ions is governed by the unitary operation (\ref{rho12t}). We can calculate the Fisher information corresponding to the parameters $T_1$ and $T_2$ via the number of excitations $\hat{\Pi}_1 = \left|k\right\rangle\left\langle k\right|$ in the longitudinal vibrational degree of freedom of ion 1.    

  For the number of phonon measurements, the Fisher information for ion 1 and ion 2 is computed as
 \begin{eqnarray} \label{Fisher}
   F_{11} &=& \sum_{k_1} P_{1}(k) \left(\frac{\partial \ln P_{1}(k)}{\partial T_1}\right)^2, \\ 
   F_{22} &=&  \sum_{k} P_{1}(k) \left(\frac{\partial \ln P_{1}(k)}{\partial T_2}\right)^2, \\  
   F_{12} &=&  F_{21} = \sum_{k}  P_{1}(k) \left(\frac{\partial \ln P_{1}(k)}{\partial T_1}\right)\left(\frac{\partial \ln P_{1}(k)}{\partial T_2}\right), 
\end{eqnarray}   
  where  $P_1(k)$ is the probability distribution of the occupation number obtained by measuring the energy of Ion 1.

\begin{figure*}[ht] 
\centering
\includegraphics[scale=0.6]{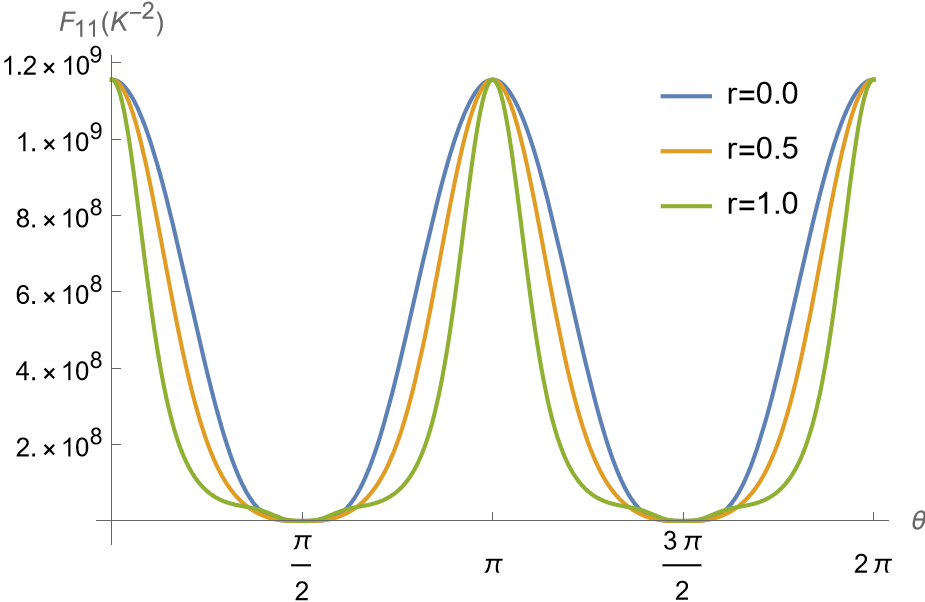} \includegraphics[scale=0.6]{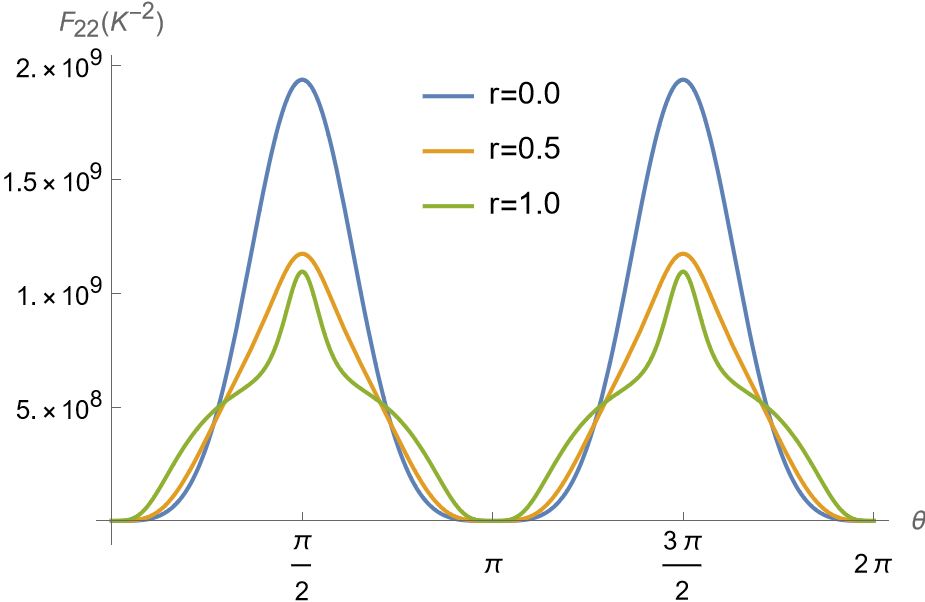} 
\includegraphics[scale=0.6]{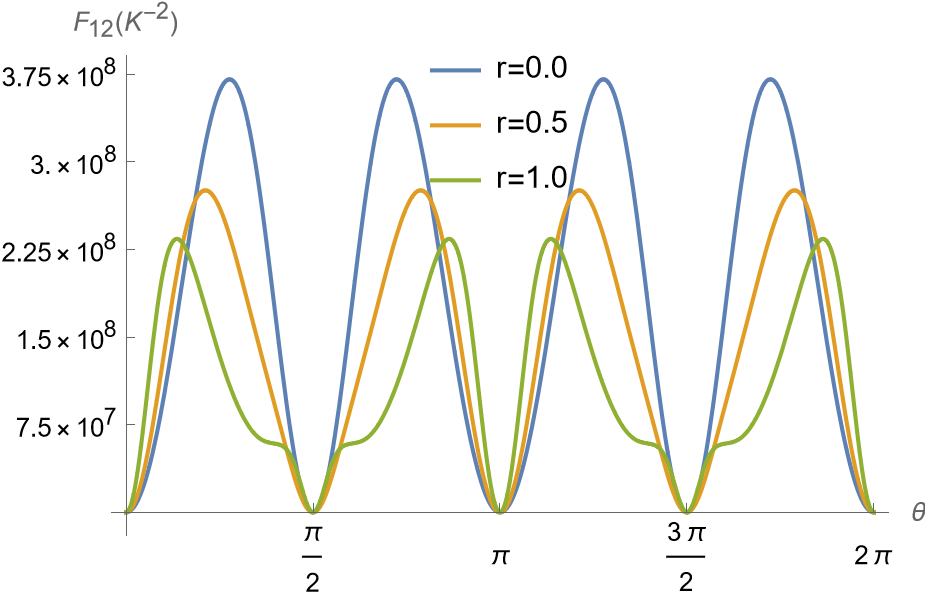} 
\caption{Fisher information  $F_{11}$, $F_{22}$ and $F_{12}$ as a function of parameter $\theta=gt$, for and $T_{1}=2.8\times10^{-5}K$, and $T_{2}=2.08\times10^{-5}K$. Here we have fixed $\Omega= 4$ MHz. \cite{seidelin2006microfabricated}.}
\label{FIs}
\end{figure*}

 \begin{figure*}[ht] 
\centering
\includegraphics[scale=0.6]{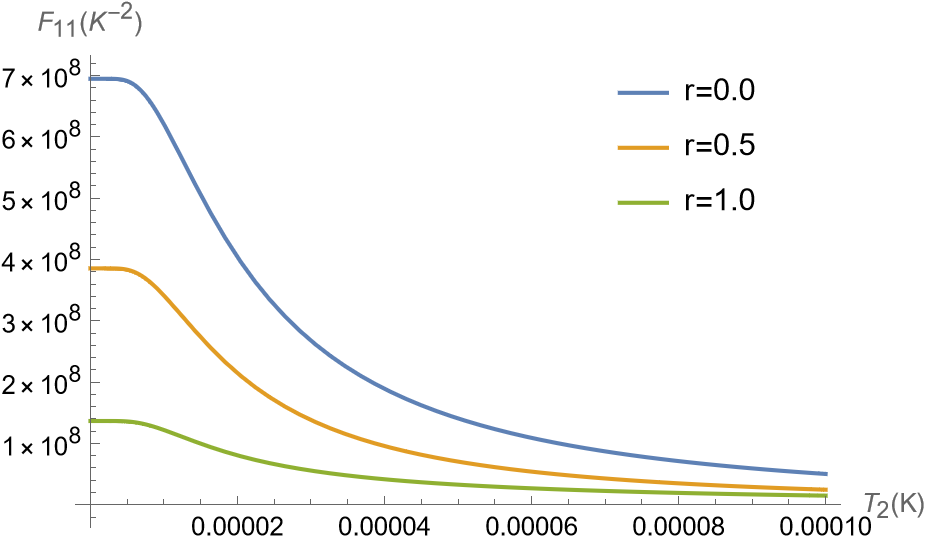} \includegraphics[scale=0.6]{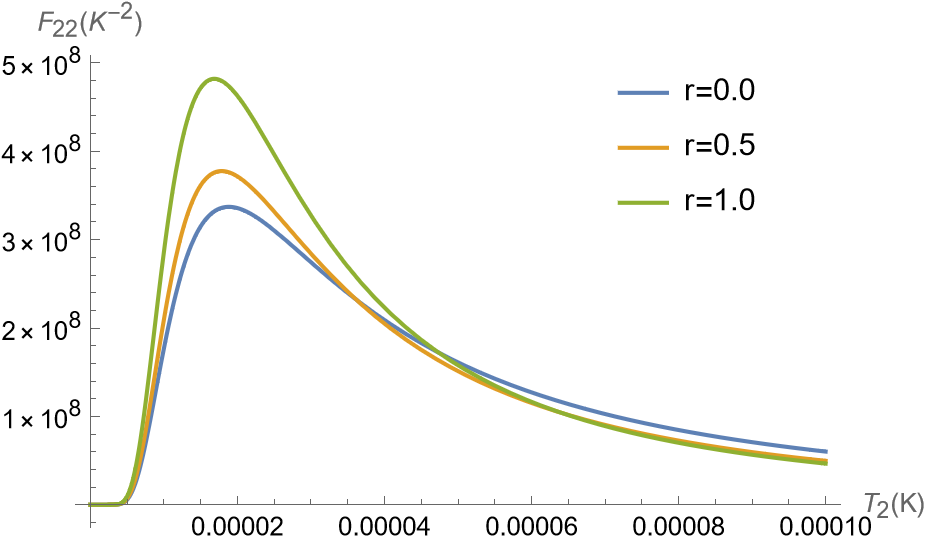} 
\includegraphics[scale=0.6]{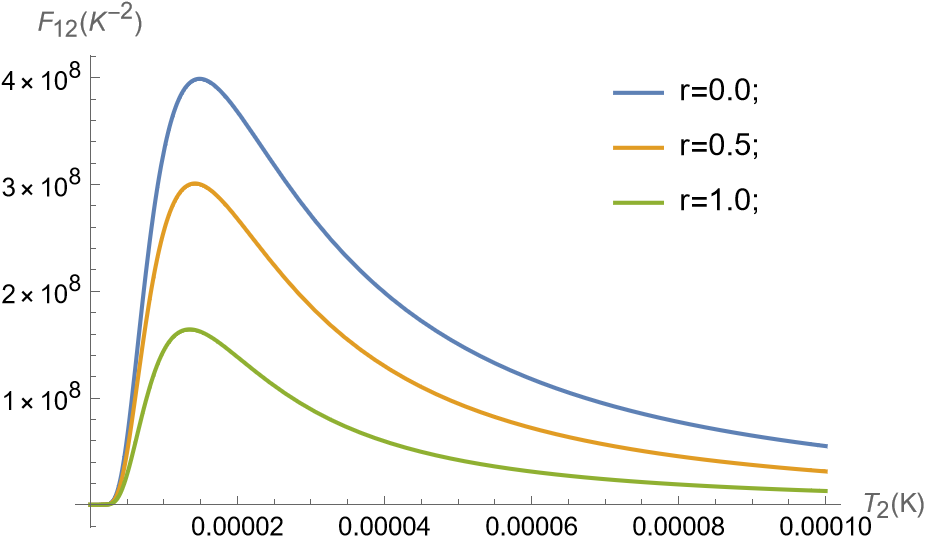} 
\caption{ Fisher information  $F_{11}$, $F_{22}$ and $F_{12}$ as a function of parameter $T_{2}$ for $\theta=\pi/4$ and $T_{1}=2.8\times10^{-5}K$. Here we have fixed $\Omega = 4$ MHz \cite{seidelin2006microfabricated}.}
\label{F1F2}
\end{figure*}

The probability distribution of $k$ phonons is given by
\begin{equation}
    P_{1}(k)=\frac{1}{k!\sqrt{\det ({\mathbf V'}-\frac{1}{2}{\mathbf I})}}\int d^{2}\gamma\,\, |\gamma|^{2k} e^{\frac{1}{2}\gamma^{\dagger} {\mathbf Z}({\mathbf V'}-\frac{1}{2}{\mathbf I}){\mathbf Z}\gamma-|\gamma|^2 } 
    \label{pk0},
\end{equation}
where
\begin{equation}
 {\mathbf V'}-\frac{1}{2}{\mathbf I}=\begin{pmatrix}
a & b \\
b & a 
\end{pmatrix},
\end{equation}
with $a=\bar{n}_{1}\cos^2 {(gt)} +\left[\left(\bar{n}_2+1/2\right)\cosh{2r}-1/2\right]$, and $b=\left( 
\bar{n}_2 +1/2\right)\sinh{(2r)} \sin^2 {(gt)}$.

The probability distribution given by Eq. \ref{pk0} 
involves a Gaussian integral that can be evaluated exactly so that the probability distribution is given as follows
\begin{equation}
 P_{1}(k)=\frac{(-1)^{k}}{k!\sqrt{a^2-b^2}}\frac{\partial^{k}}{\partial A^{k}}  \left( \frac{1}{\sqrt{A^2-B^2}}\right)
 \label{pk1},
\end{equation}
where
\begin{equation}
    A^2=\left(1+\frac{a}{a^2-b^2}\right),
\end{equation}
and
\begin{equation}
    B^2=\frac{b^2}{a^2-b^2}.
\end{equation}

Now using the series expansion
\begin{equation}
    \frac{1}{\sqrt{A^2-B^2}}=\frac{1}{A}\sum_{n=0}^{\infty }\frac{(2n)!}{2^{2n}(n!)^2}\left(\frac{B}{A}
    \right)^{2n},
\end{equation}
the derivatives in relation to $A$ are easily evaluated, so the probability distribution reduces to
\begin{equation}
    P_{1}(k)=\frac{1}{A^{k+1}\, \sqrt{a^2-b^2}}\sum_{n=0}^{\infty}\frac{(2n+k)!}{k!(2^{n}n!)^2}\left(\frac{B}{A}
    \right)^{2n}
    \label{pk2},
\end{equation}
The series in Eq. \ref{pk2} can be identified with the Hypergeometric function $ \mathstrut_{2}
F_{1} \left(\frac{1+k}{2},\frac{2+k}{2},1,\left(\frac{B}{A}\right)^2\right)$ \cite{arfken} so we can write
\begin{equation}
   P_{1}(k)= \frac{ \mathstrut_2 F_{1} \left(\frac{1+k}{2},\frac{2+k}{2},1,\left(\frac{B}{A}\right)^2\right)}{A^{k+1}\sqrt{a^2-b^2}}
   \label{pk3}.
\end{equation}

In the case that $r=0$, with both modes in thermal states, the probability distribution reduces to the simple expression
\begin{equation}
 P_{1}(k) = \frac{\left[\overline{n}_{1}\cos^2{(gt)}+ \overline{n}_{2}\sin^2{(gt)}\right]^k}{\left[1+\overline{n}_{1}\cos^2{(gt)}+ \overline{n}_{2}\sin^2{(gt)}\right]^{k+1}}.
\end{equation}

For the calculations of the elements of the Fisher information we set $k=200$  in order to perform the numerical calculations.
  
  In Fig. \ref{FIs}, the Fisher information  $F_{11}$, $F_{22}$ and $F_{12}$ evolution with $\theta=gt$, is represented for two thermal reservoirs (for $r=0$) at $T_{1}=2.8\times10^{-5}K$, and $T_{2}=2.08\times10^{-5}K$ and for a non-thermal reservoir (for $r\neq 0$). We can see by the corresponding curves of $F_{11}$ that the temperature $T_1$ is best inferred at the instants where $\theta=n\pi$, for $n=0,1,2,...$. This occurs because at those instants the initial equilibrium state of the ion 1 has recurred\cite{Oliveira_1999}. In contrast, as can be seen in $F_{22}$, at $\theta= m \frac {\pi}{2}$, for $m= 1, 3, 5,...$ the state of ions 1 and 2 has been swapped, by the characteristic of evolution (\ref{ucor}), and therefore is not surprising that those are the best instants for inference of temperature of the vibrational mode of ion 2 -- it occurs because when the states are interchanged\cite{Oliveira_1999}, when detection of the ion 1 vibrational mode is performed, in fact, we are effectively measuring the state of ion 2. Simultaneous inference of both ions temperatures occurs when $\theta= m \frac {\pi}{4}$, for $m= 1, 3, 5,...$, as depicted by $F_{12}$. Even when a non-thermal state is involved, the approach is successful for the inference of the ions temperature, as we can see in the same figures by varying $r$, the same characteristics hold. The effect of squeezing, however is to improve the entanglement and therefore it turns out that the distribution of points where the inference is optimal is spreaded, meaning that there is an advantage in using entanglement as a way to improve the accessibility at distinct instants of time for inference.

\begin{figure*}[ht] 
\centering
\includegraphics[scale=0.6]{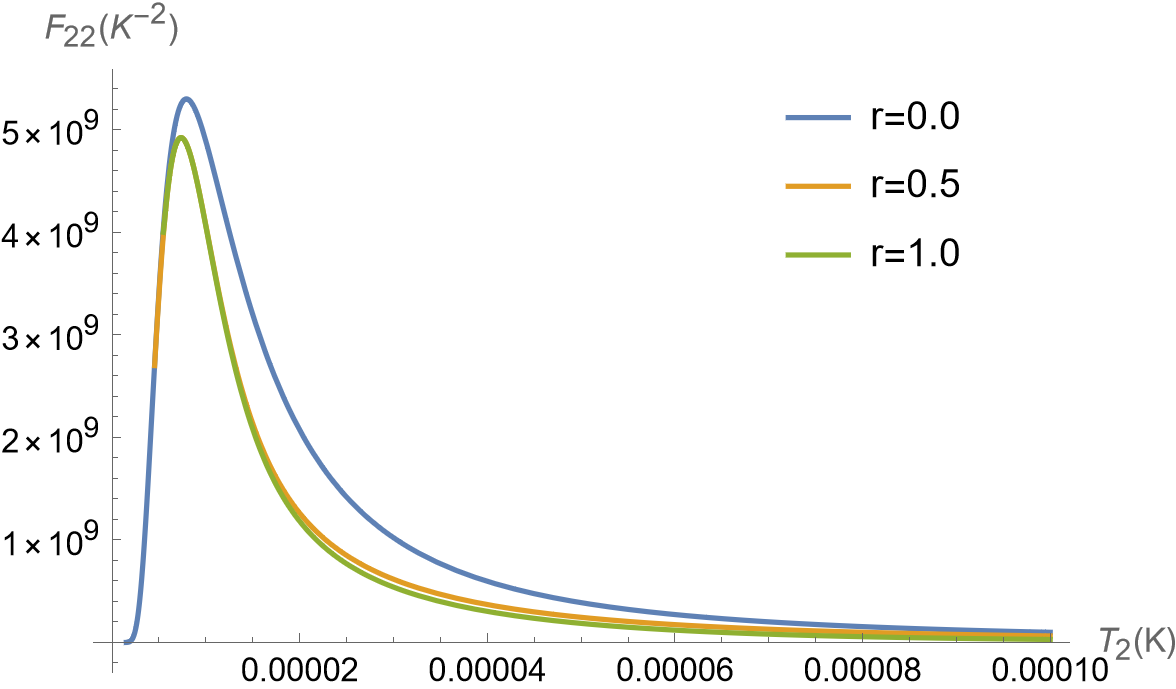}
\caption{Fisher information  $F_{22}$ as a function of parameter $T_{2}$ for $\theta=\pi/2$ and $T_{1}=2.8\times10^{-5}K$. Here we have fixed $\Omega= 4$ MHz \cite{seidelin2006microfabricated}. At $\theta=\pi/2$, both 
$F_{11}$ and $F_{12}$ are null for any $T_2$.}
\label{Fipi2}
\end{figure*}

   In Fig. (\ref{F1F2}), the Fisher information $F_{11}$, $F_{22}$ and $F_{12}$ are depicted as functions of the temperature $T_{2}$ for $\theta=\pi/4$ (similar results are observed for odd multiples of $\theta=\pi/4$), and  different values of $r$, for a fixed temperature of ion 1, $T_{1}=2.8\times10^{-5}K$.  This result indicates a precision loss for the estimation of temperature of the 1-ion, when we increase the $r$ (this is reproduced in the graphs of $F_{11}$ and $F{12}$ in Fig. (\ref{F1F2})).
The same cannot be said for estimating only the temperature $T_{2}$, which improves with an increase in $r$ (see graphs of $F_{22}$ in Fig. (\ref{F1F2})) under the same conditions, working as a super probe in the non-classical regime $2(\bar{n}_{2}+1/2)\leq e^{2r}$, for this maximally entangled situation for $\theta=\pi/4$. In fact, the  maximal precision is still observed at $\theta=\pi/2$, as depicted in Fig. \ref{Fipi2}, in contrast to $F_{11}$ and $F_{12}$, which are null for any $T_2$ we see that $F_{22}$ reaches optimal values, which however is disturbed by the non-thermal reservoir feature.

\section{Concluding Remarks} 

In summary, we have considered a simple electrostatic interaction architecture between two trapped ions, in which this interaction is approximated to a beam-splitter interaction. For Entanglement analysis and temperature estimation, we take ion 1 with a thermal initial state, and ion 2 with the initial state to be thermal and non-thermal (due to quadrature squeezing). We use the interaction to effectively infer the temperature of the individual ions, by optimising the interaction  time evolution. It was observed that the estimation of the temperature of ion 2 is improved when the initial state is non-classical. Simultaneous measurements are often difficult to implement, in particular when associated with motional degrees of ions. The most practical way is to estimate the temperature by measuring each ion. Here, we chose to measure the ion that does not contain a possible squeezing, making it as a classical channel probe, contrary to what it has in the literature, where they use detectors with quantum properties for estimating parameters of classical systems \cite{latune2013quantum, campbell2017global, caves1980quantum, aasi2013enhanced}.

\section*{Acknowledgements}

HASC wishes to acknowledge the financial support from Brazilian funding agency CAPES. This work was partially supported by the ''EDITAL FAPEPI/MCT/CNPq N$^{o}$ 007/2018: Programa de Infraestrutura para Jovens Pesquisadores/Programa Primeiro Projetos (PPP)'', and CNPq.

\bibliography{fisher}



\end{document}